\documentclass[aps,prd, superscriptaddress,showpacs,nofootinbib]{revtex4}
\usepackage{amsmath,amssymb}

\usepackage{hyperref}
\usepackage{graphicx}

\newcommand{\be}{\begin{eqnarray}}
\newcommand{\ee}{\end{eqnarray}}
\newcommand{\ba}{\begin{eqnarray}}
\newcommand{\ea}{\end{eqnarray}}

\def\ben{\begin{equation}}
\def\een{\end{equation}}
\def\bena{\begin{eqnarray}}
\def\eena{\end{eqnarray}}

\renewcommand{\d}{\mbox{{\rm d}}}

\begin{document}

\begin{flushright}{\scriptsize LMU-ASC 36/11 \\  \scriptsize ROM2F/2011/10 }
\end{flushright}
\title{Introducing the Slotheon: a slow Galileon scalar field in curved space-time}

\author{Cristiano~Germani}
\email{cristiano.germani@lmu.de}
\affiliation{Arnold Sommerfeld Center, Ludwig-Maximilians-University, Theresienstr. 37, 80333 Muenchen, Germany}

\author{Luca~Martucci}
\email{luca.martucci@roma2.infn.it}
\affiliation{I.N.F.N. Sezione di Roma ``Tor Vergata'' \&  Dipartimento di Fisica, Universit\`a di Roma ``TorVergata",
Via della Ricerca ScientiÞca, 00133 Roma, Italy}

\author{Parvin~Moyassari}
\email{parvin.moyassari@physik.lmu.de}
\affiliation{Excellence Cluster Universe, Boltzmannstr.~2, 85748 Garching, Germany}
\affiliation{Arnold Sommerfeld Center, Ludwig-Maximilians-University, Theresienstr. 37, 80333 Muenchen, Germany}

\begin{abstract}
%cr
In this paper, we define covariant Galilean transformations in curved spacetime and find all scalar field theories invariant under this symmetry.  
The Slotheon is a Galilean invariant scalar field with a modified propagator such that, whenever gravity is turned on and energy conditions are not violated, it moves ``slower''
than in the canonical set-up. This property is achieved by a non-minimal derivative coupling of the Slotheon to the Einstein tensor. 
We prove that spherically symmetric black holes cannot have Slotheonic hairs. We then notice that in small derivative regimes the theory has an asymptotic {\it local} shift symmetry whenever the non-canonical coupling dominates over the canonical one.

\end{abstract}

\maketitle

\section{Introduction}

\noindent Undoubtedly, the search for theories with special symmetries is a key issue in theoretical physics. Indeed, usually, such theories have the advantage of being quantum mechanically under control. 

One of the simplest possible symmetries is the shift invariance of a scalar field $\pi$, i.e. the symmetry under the shift
\be\label{trivial}
\pi\rightarrow\pi+c\ ,
\ee
where $c$ is a constant. However, such a symmetry is not very interesting as any theory involving only derivatives of $\pi$ would be invariant under (\ref{trivial}). The question is then whether such symmetry may be generalized  to a more complicated shift
\be\label{constr}
\pi\rightarrow\pi+f(x)\ ,
\ee
where $f(x)$ is some specific function of space-time coordinates, depending on some constants parametrizing  the independent symmetries encoded in (\ref{constr}). The class of Lagrangians invariant under (\ref{constr}) will be more and more constrained, depending on the degree of arbitrariness of $f(x)$. The extreme case is the one in which $f(x)$ is a completely arbitrary function and then (\ref{constr}) can be regarded as  a gauge symmetry.

%The question is then whether such symmetry may be ``gauged'' by a more complicated shift
%\be\label{constr}
%\pi\rightarrow\pi+f(x)\ ,
%\ee
%where $f(x)$ is some specific function of space-time coordinates. In this case the class of Lagrangians invariant under (\ref{constr}) may be highly constrained. The extreme case is in which $f(x)$ is any arbitrary function. There the scalar field $\pi$ cannot be anymore a propagating degree of freedom. 

The next to trivial shift symmetry is what is commonly called Galileon shift \cite{galilean}. This symmetry, formulated solely in flat (Minkowski) space, is an on-shell symmetry. In other words, the equation of motion are invariant under the Galileon shift
\be\label{flat}
\pi\rightarrow \pi+c+c_\mu x^\mu\ ,
\ee
with $c$ and $c_\mu$ respectively a constant and a constant vector, whereas the action shifts by a total derivative which gives a {\it non-vanishing} boundary contribution. Mainly inspired by the decoupling limit of the Dvali-Gabadadze-Porrati (DGP) model \cite{dgp}, the Authors of \cite{galilean} showed that in flat space, there exist only four forms of scalar field Lagrangians with second order field equations and invariant under the Galileon symmetry. These theories, turned out to admit a non-renormalization theorem. In other words, it is proven that the mass parameters in the Galileon terms do not get renormalized \cite{non,non2}\footnote{Note however that any other operator can be generated by loops, see for example \cite{nr,referee}.}. 

In a subsequent analysis, the Authors of \cite{covariant} showed that healthy covariantization of the flat space Galileon invariant theories, would generically break the flat space Galileon invariance. This is mainly due to the fact that the constant form $c_\mu$, is not shear free, i.e. $\nabla_{(\alpha} c_{\beta)}\neq 0$, where we defined $v_{(\alpha\beta)}=\frac{1}{2}\left(v_{\alpha\beta}+v_{\beta\alpha}\right)$.
Indeed, by inserting the transformation (\ref{flat}) in the equation of motion for the scalar field one always get terms proportional to the shear of $c_\alpha$\footnote{Note that since we like to obtain a scalar equation, only shear and not vorticity (the antisymmetric part) of the covariant derivative of $c_\alpha$ enters in the shifted equations.}.

The question is then whether any symmetric scalar field theory under the shift (\ref{constr}) can be constructed in a {\it fixed} curved space-time. Moreover, we will {\it always} implicitly consider only theories with equations of motion (for both gravity and scalar field) that are up to second order, although we will not explicitly state it anymore. 

In a manifold with covariantly constant Killing vectors, a Galileon symmetry similar to (\ref{flat}), may indeed be realized. In this case however, as we shall prove it, only the following Lagrangians can be constructed in contrast to the flat space case. They are \footnote{Note that in addition there are also ``tadpole" terms such as
$$
{\cal L}_{\rm tp}=M_{tp}^3\pi\left(1+\frac{R}{\mu_1^2}+\frac{GB}{\mu_2^4}\right)\ ,
$$
where $M_{tp}$ and $\mu_{1,2}$ are mass scales and $GB$ is the Gauss-Bonnet term. We will not discuss them here as we focus on source free equation of motion for $\pi$. Nevertheless, even considering them, they will not be invariant under the approximate gauged shift symmetry (\ref{constr}).}
%cr
\be\label{K}
{\cal L}_{2}={\cal L}_{2}^{\text{m}}+ {\cal L}_{2}^{\text{nm}}\equiv-\frac{1}{2}g^{\mu\nu}\partial_\mu\pi\partial_\nu\pi+\frac{1}{2 M_2^2}G^{\mu\nu}\partial_\mu\pi\partial_\nu\pi\ ,
\ee
\be\label{K1}
{\cal L}_{3}={\cal L}_{3}^{\text{m}}+ {\cal L}_{3}^{\text{nm}}\equiv\pm\frac{1}{2M_3^3}(\partial\pi)^2\square\pi\pm\frac{1}{2M_5^5} {}^{**}R^{\alpha\beta\mu\nu}\partial_\alpha\pi\partial_\mu\pi\nabla_\beta\nabla_\nu\pi\ .
\ee
where $M_i$ are mass scales for the operators of dimension $i+4$. $G_{\mu\nu}$ and ${}^{**}R^{\alpha\beta\mu\nu}$   are respectively the Einstein and double dual Riemann tensors \footnote{The double dual Riemann tensor is defined as $${}^{**}R^{\mu_1\mu_2\nu_1\nu_2}\equiv-\frac{1}{4}\mathcal{E}^{\mu_1 \mu_2\mu_3\mu_4}~
\mathcal{E}^{\nu_1 \nu_2\nu_3\nu_4}R_{\mu_3\mu_4\nu_3 \nu_4}\ ,$$ where $$ \mathcal{E}^{\mu_1 \mu_2\mu_3\mu_4}=-\frac{1}{\sqrt{-g}}\delta_1^{[\mu_1}\delta_2^{\mu_2}\delta_3^{\mu_3}\delta_4^{\mu_4]}\ .$$}. Note that, at this level, only the quadratic canonical Lagrangian may have a definite sign in order to avoid ghost propagation around Minkowski. Instead, the sign of ${\cal L}_2^{\rm nm}$ is chosen in such a way to avoid ghosts when the weak energy condition $G^{tt}\geq 0$ is satisfied \footnote{We would like to stress that this condition is not enough to guarantee the absence of ghost propagation whenever the metric is dynamical.}.

Next, one can remove the requirement of the existence of covariantly conserved Killing vector and couple the theory (\ref{K}) to a dynamical metric, by adding the standard Einstein-Hilbert term and possibly also a potential for $\pi$. In this case, we will show that for parity invariant Lagrangians ($\pi\rightarrow -\pi$), in the small derivatives regime of the scalar, an approximate infinitesimal shift symmetry (\ref{constr}) emerges for the theory  ${\cal L}_{2}^{\text{nm}}$, if and only if an appropriate shift of the metric is also considered. The theory ${\cal L}_{2}^{\text{nm}}$, in this regime, is the base for the Gravitationally-Enhanced-Friction (GEF) models of inflation \cite{yuki, new, uv}. Therefore, thanks to the additional symmetry, the GEF models are endowed, during inflation (small scalar field derivatives) with a protection against quantum corrections to the effective Lagrangian up to the Planck scales, if the potential terms only softly break the gauged shift symmetry (\ref{constr}). In fact, although other terms may be generated by loops that are invariant under the symmetry (\ref{constr}), they will be generically suppressed by either slow roll or higher powers of Planck mass. This is mainly due to the fact that in this theory, the gravity strong coupling is still at the Planck scale, as we shall show.

By adding the standard Einstein-Hilbert term to ${\cal L}_{2}$, and possibly a non trivial potential for $\pi$, one gets a simple though rich gravitational theory, with some nice peculiarities. In particular, in regimes in which the analogue of the weak energy condition is valid, the field $\pi$ moves `slower' than in the cousin canonical theory. For this reason,  we dub $\pi$ as the  {\em Slotheon} and the Slotheonic nature of this theory is, in fact, at the origin of the efficiency of the GEF models.

Furthermore, we show that the Slotheonic theory has only spherically symmetric black hole solutions with no scalar hairs and we find indications that this property should hold for any black hole solutions. This important result combined with previous analysis in homogeneous and isotropic space-times \cite{yuki}, is a step forward to prove the stability of this theory.

\section{Shift and accidental symmetries}
\label{secII}
%cr
\noindent In this section we would like to find all the possible scalar field Lagrangians (collectively denoted as ${\cal L}$), that do not contain purely potential terms, according to some symmetry principle. %we focus on Lagrangians that do not contain purely potential terms  for the scalar. 

First of all, being $\pi$ a real scalar, it is natural to impose on ${\cal L}$ the shift symmetry (\ref{trivial}).
In other words, we assume that ${\cal L}$  depends only on derivatives of $\pi$.  
Furthermore, in order to avoid problems with ghosts, due to the so called Ostrogradski instability \cite{ghost} we also impose that the equations of motion contain at most second order derivatives.

These requirements still leaves a large number of possible ${\cal L}$. In order to further restricts the form of the action we  require that,
for certain background metrics, the shift symmetry (\ref{trivial}) {\em enhances} to accidental {\em point-dependent} shift-symmetries of the form (\ref{constr}). 

The prototype example of such an effect is provided by the Galileon symmetry for {\em flat} space-time \cite{galilean}, in which (\ref{trivial}) enhances to (\ref{flat}). We would like to preserve the Galileon symmetry for the case of flat space-time and generalized it to more general, though still restricted, curved space-times. 

As we shall discuss, the on-shell flat space-time symmetry (\ref{flat}) can be naturally generalized in background metrics with a certain number of covariantly constant vectors $\xi_a=\xi_a^\mu\partial_\mu$.
Requiring that the equations of motion derived from ${\cal L}$ preserve such accidental symmetries will lead us to consider only the set ${\cal L}_{2,3}$.

\subsection{Galileon symmetry in curved space-time}\label{sec:cG}

In flat spacetime, the transformations (\ref{flat}) can be written in a covariant way by introducing the translation Killing vectors of Minkowski, i.e. $\xi_a^\mu=\delta_a^\mu$ in Cartesian coordinates. These Killing vectors are very special as they have the properties of being convariantly constant, i.e. $\nabla_\mu\xi_a=0$.

Our definition of Galilean transformation in curved spacetime will just be the straightforward generalization of the flat case. Let us introduce the one forms $\xi^a=\xi^a_\mu\d x^\mu$ dual to the Killing vectors: $\xi^a_\mu=g_{\mu\nu}\xi^\nu_a$. Hence, we will require that our theories are invariant under
\be\label{covdifshift}
\pi_\mu\rightarrow \pi_\mu+c_a\,\xi^a_\mu\ ,
\ee
where $c_a$ are constants, latin indices are contracted with Euclidean metric and
we have used  the notation
\be
\pi_{\mu_1\ldots\mu_k}\equiv \nabla_{\mu_1}\ldots\nabla_{\mu_k}\pi\ .
\ee
Consistency of (\ref{covdifshift}) requires the one forms $\xi^a$ to be closed:
\be\label{intcond}
\d\xi^a=0\ .
\ee 
In other words,  $\xi_a$ must be covariantly constant and we will loosely say that the Killing vectors must be integrable \footnote{Note, that one could find different symmetries. For example, if one relaxes the constant shift invariance (\ref{trivial}) in curved space-time, one can find other theories invariant under specific shifts $\pi\rightarrow \pi+c(x)$, where $c(x)$ is a function of curvatures. A typical example is given in \cite{burrage,goon1}. We thank Claudia de Rham for pointing this out.}.  

Space-times admitting integrable Killing vectors are of particular type \cite{Stefani}.  A  Killing vector $\xi^\mu$ can be covariantly constant only if $\xi$ satisfies the algebraic condition  
\be\label{ruin}
R^{\mu}{}_{\nu\rho\sigma}\xi^\nu=0\ ,
\ee
which can be obtained from the consistency condition $[\nabla_\rho,\nabla_\sigma]\xi^\mu=0$.  In other words, the holonomy group of space-time must be reduced to a subgroup of SO$(1,3)$. Explicitly, if the 
vector is non-null, the space-time metric is of the form
\be
\d s^2=g_{ij}(x^k) \d x^i \d x^j +\kappa\, \d y^2\, , ~~~i,j,k=1,2,3\ ,
\ee
where $\kappa=+1,-1$ for spacelike or timelike $\xi^\mu$, respectively, or for a null $\xi^\mu$
\be
\d s^2=g_{ij}(x^k) \d x^i \d x^j +\d z \d y\, , ~~~i,j,k=1,2,3\ ,
\ee
where $z$ is any coordinate in the $i$'s directions.

Given a set of integrable Killing vectors $\xi_a$ we can easily integrate (\ref{covdifshift}) into a curved Galileon transformation
\be\label{curvedG}
\pi(x)\rightarrow \pi(x)+c+c_a\int^x_{\gamma,x_0} \xi^a,
\ee
where we have chosen a certain reference point $x_0$ and a curve $\gamma$ connecting $x$ with $x_0$. Thanks to (\ref{intcond}), this quantity is well defined. Indeed,  it does not change under continuous deformation of the curve $\gamma$. Furthermore, the change of the reference point $x_0$ can be reabsorbed into a shift of $c$. 

The transformation (\ref{curvedG}) represents our proposal of curved Galileon symmetry. Let us revisit the Minkowski case in this covariant language.  In that case, the integrable Killing vectors are the four generators of the translations. Fixing $x^\mu$ to be the Minkowskian coordinates, the associated one-forms take the form 
\be
\xi^a\Big|_{\rm Mink_4}=\delta^a_\mu \d x^\mu\ .
\ee
By choosing $x_0$ as the origin $x^\mu=0$ it is immediate to see that (\ref{curvedG}) reproduces (\ref{flat}), with $c_\mu\equiv c_a\delta^a_\mu$.

\subsection{Galileon invariant theories in curved space-time}\label{sec}

We would like now to find a Galilean invariant theory in curved spacetime where the metric is {\it non-dynamical}. Later on we will drop this last requirement. The Lagrangian defining the theory we look for is either 1) the covariantized version of the Galileon Lagrangian in Minkowski or 2) made of terms which vanish once restricted to flat space. 
 
We start with theories that are not trivial once restricted to flat space, i.e. the case 1). In flat space, Galilean invariant theories were classified by \cite{galilean}. To keep the equation of motion second order in curved spacetime, the Authors in \cite{covariant} showed that the original flat space self couplings of the scalar field derivatives must be supplemented by non-minimal couplings to curvatures.

We can now check directly what theories among the one classified in \cite{covariant} are invariant under the Galilean symmetries in curved spacetime (\ref{covdifshift}).  The key point to bare in mind is that the shift of the scalar field derivative is covariantly constant and, the same shift, once contracted to curvatures vanishes as in (\ref{ruin}). Therefore, if the scalar field equation of motion contains terms proportional to one derivative of $\pi$ without contraction to curvatures, then, the theory is {\it not} Galilean invariant in curved spacetime. An example of those terms is $R(\partial\pi)^2$.

Inspecting the set of four Lagrangians found in \cite{covariant}, the only Galilean invariant theories in curved spacetime which are not trivial in the flat limit are ${\cal L}_2^{\rm m},{\cal L}_3^{\rm m}$.

We now consider theories that are solely non-minimally coupled, i.e. they vanish in the flat limit. These theories, and in fact all possible scalar-tensor theories with second order equation of motion, are classified by Hordenski in \cite{hord}. We can then easily check that the only theories invariant under the Galilean symmetry in curved spacetime are ${\cal L}_2^{\rm nm}, {\cal L}_3^{\rm nm}$. This is again due to the fact that terms proportional to one derivative of $\pi$ without contraction to curvatures are not Galilean invariant in curved spacetime.

In conclusion, the only invariant theory under Galilean symmetry in curved spacetime (\ref{covdifshift}), parameterized by the masses $M_i$, is
\be\label{L2}
S=-\frac{1}{2}\int \d^4x\sqrt{-g}\left({\cal L}_2+{\cal L}_3\right)\ .
\ee

It is interesting to note that the theory (\ref{L2}), as in the flat case \cite{non}, follows a non-renormalization theorem for the mass parameters $M_i$, whenever gravity is non dynamical. This is due to the fact that in each cubic vertex, the algebraic structure of derivatives is not different from the one of the flat space, thanks to Bianchi identities. In other words, one can easily check that cubic interactions only produce effective higher-derivatives operators and therefore cannot renormalize $M_i$. Specifically, one sees, following \cite{non} for the flat case, that vertexes of type
\be
\sqrt{-g}{}^{**}R^{\alpha\beta\mu\nu}\partial_\alpha\pi_{\rm ext}\partial_\mu\pi_{\rm int}\nabla_{\beta\nu}\pi_{\rm int}=-\sqrt{-g}{}^{**}R^{\alpha\beta\mu\nu}\nabla_{\alpha\nu}\pi_{\rm ext}\partial_\mu\pi_{\rm int}\partial_{\beta}\pi_{\rm int}+ {\rm boundaries}\ ,
\ee
where $\pi_{\rm ext}$ and $\pi_{\rm int}$ are respectively the external and the internal legs of a diagram involving loops, cannot renormalize $M_3$ as they are equivalent to higher-derivatives powers in $\pi_{\rm ext}$.

Obviously, as the quadratic Lagrangian has no any interactions in the case of non-dynamical metric, $M_2$ is not renormalized as well. 

This conclusion would change in the case in which gravity is dynamical. Nevertheless, in this case, the parameters $M_i$ would only have runnings suppressed by $M_{\rm P}$, as we shall discuss later on. 

\section{Gauging the shift invariance in curved space-time}

\noindent As we showed before, only few manifolds may support a Galileon symmetry. We may however ask whether the Galileon invariance introduced earlier can be recovered, even in some approximate sense, in space-times with no integrable Killing vectors. 

An obvious requirement is that, if such an approximate symmetry exists, it should be also realized point-wise. Locally indeed we can always define Riemann coordinates $(x_R^\mu)$ around any point $P$ such that, for any constant form $c_{\mu}$ in this coordinates
\be
\nabla_\mu c_\nu={\cal O}(x_R)\ .
\ee
In this case we can ask whether there exists any theory invariant under the local Galileon symmetry
\be\label{27}
\pi\rightarrow\pi+c+c_\mu x_R^\mu\ ,
\ee
up to order ${\cal O}(x_R)$. Note that, although the Christoffel symbols vanish up to order ${\cal O}(x_R)$, curvatures do not vanish. Let us then consider the scalar field equations of motion of theories (\ref{K}) and (\ref{K1})
\be\label{eqm}
E_2:&=&\left(g^{\mu\nu}-\frac{G^{\mu\nu}}{M^2_2}\right)\pi_{\mu\nu}\ ,\cr\nonumber
E_3:&=&\pm\left(\pm\frac{ M_3^3}{M_5^5}{}^{**}R^{\mu_1\mu_2\nu_1\nu_2}+g^{\mu_1\nu_1}g^{\mu_2\nu_2}-g^{\mu_1\nu_2}g^{\mu_2\nu_1}\right)\left(\pi_{\mu_1\nu_1}\pi_{\mu_2\nu_2}+R^\alpha{}_{\mu_1\nu_1\mu_2}\pi_\alpha\pi_{\nu_2}\right)\ .
\ee
We can easily see that only $E_2$ is invariant under the approximate shift (\ref{27}) up to distance $x_R\sim \ell$ where $\ell$ is the local curvature radius of the spacetime. 

At this level however, the approximate symmetry (\ref{27}) cannot be extended far away from the point $P$. It is then clear that gravity should participate to the Galileon shift in order to extend this symmetry at distances such that the Christoffel symbols cannot be neglected. 
We will then only focus on the theory ${\cal L}_2$. This theory can be singled out by requiring the action to be invariant under the following additional discrete symmetry
\be
\pi\rightarrow-\pi\ ,
\ee
which we will loosely call $\pi$-parity.

Let us then study the following action: 
\be\label{jordan}
S(g,\pi)=\int \d^4 x \sqrt{-g} \left[\frac{M_{\rm P}^2}{2} R(g)+{\cal L}_2\right]\ .
\ee
Now we can make the metric $g_{\mu\nu}$ participate actively, enlarging the possibility of identifying the relevant (approximate) symmetries of the form (\ref{constr}). 

Consider the following small derivative expansion regime
\be\label{regime}
\varepsilon\sim \frac{(\partial\pi)^2}{M_2^2 M^2_{\rm P}}\ll 1\ ,
\ee
and note that \cite{fab}
\be
\int d^4x\sqrt{-g} G^{\alpha\beta} \pi_\alpha\pi_\beta&=&M_2^2 M^2_{\rm P}\int d^4x \frac{\delta{\sqrt{-g}R}}{\delta g_{\alpha\beta}}\delta g_{\alpha\beta}\Big |_{\delta g_{\alpha\beta}=-\frac{\partial_\alpha\pi\partial_\beta\pi}{M_2^2M_{\rm P}^2 }}+{\rm boundaries}\ .
\ee
We find that (\ref{jordan}) can be found as an expansion of ${\cal O} (\varepsilon^2)$ of the following action \footnote{Our quadratic action (\ref{einstein}) agrees with \cite{kurt} and disagrees with \cite{claudia}. This can be seen by noticing that the purely derivative quadratic terms in $\pi$ of \cite{kurt} (Eq. (31) of the cited paper) is nothing else than the Ricci scalar coupled to the kinetic term of $\pi$ plus boundary terms. The disagreement is due to a missing factor upon passing from the correct expansion Eq.(77)  to the Lagrangian Eq.(79) of \cite{claudia}.}
\be\label{einstein}
\hat S(h,\pi)=\frac{1}{2}\int \d^4x\, \sqrt{-h} \left[M_{\rm P}^2 R(h)-h^{\mu\nu}\partial_\mu\pi\partial_\nu\pi\right]\ ,
\ee
where
\be\label{finsler}
h_{\mu\nu}\equiv g_{\mu\nu}-\frac{\partial_\mu\pi\partial_\nu\pi}{M_2^2M_{\rm P}^2 }\ ,
\ee
and $h^{\mu\nu}$ is the inverse of $h_{\mu\nu}$ which is also known as Finsler metric. Explicitly, we have
\be\label{eqactions}
\hat S(h,\pi)=S(g,\pi)+{\cal O}\left( \varepsilon^2\right)\ .
\ee
Notice that, in the regime (\ref{regime}) and to leading order in the perturbative $\varepsilon$-expansion, the canonical action $\hat S(h,\pi)$ can be regarded as the Einstein frame action of the theory (\ref{jordan}). Clearly, this is not true if (\ref{regime}) is violated, as the two theories are substantially different.

The metric $h_{\mu\nu}$ is exactly invariant if we consider the combined transformation 
\be\label{exact}
\pi\rightarrow\pi+f(x)\ ,\quad  g_{\mu\nu}\rightarrow g_{\mu\nu}+2\frac{\partial_{(\mu} f\partial_{\nu)}\pi}{M_2^2 M_{\rm P}^2}\ ,
\ee 
where we can assume that
\be\label{fregime}
\frac{\partial f}{M_2 M_{\rm P}}\sim {\cal O}(\sqrt{\varepsilon})\ .
\ee
In this way, the transformed $\pi$ continues to satisfy the small derivative condition (\ref{regime}). This simple observation has an immediate consequence in a regime in which the Lagrangian $G^{\mu\nu}\pi_\mu\pi_\nu/M_2^2$ dominates over the canonical kinetic term in (\ref{jordan}), i.e.
\be\label{HFregime}
\frac{{\cal L}_2^{\rm nm}}{{\cal L}_2^{\rm m}}\gg 1\ .
\ee
We call this the {\em high friction} regime for reasons which will become clear in the following discussions. 

If the system is in high friction regime the theory (\ref{jordan}) can be recast as a first order expansion of the Einstein-Hilbert action for the metric $h$
\be\label{EH}
S_{\rm EH}(h)=\frac{M_{\rm P}^2 }{2}\int \d^4x \sqrt{-h}\, R(h)\ .
\ee
It is now easy to see that, in the small derivative high-friction regime defined by (\ref{regime}) and (\ref{HFregime}), the Slotheon action $S(g,\pi)$ is invariant under the transformation (\ref{exact})  up to terms of order ${\cal O}\left( \varepsilon^2\right)$. 
Hence, we conclude that  in this regime the action (\ref{jordan}) has an approximate  symmetry (\ref{exact}) which `gauges'  the constant shift symmetry $\pi\rightarrow \pi+c$ by mixing $\pi$ and metric degrees of freedom. As it is clear from the above discussion, this gauge symmetry simply removes the physical degrees of freedom encoded in $\pi$, which recombines with $g$ into 
the physical Einstein metric $h$, at least to first order in $\varepsilon$.

It is interesting to compare this symmetry with the curved Galileon symmetry  discussed in section \ref{secII}. Consider a certain metric $g$ with a certain set of Killing vectors $\xi_a$ and take $f(x)$ to have the form given by (\ref{curvedG}), i.e.
\be\label{ggf}
f(x)=c+c_a\int^x_{x_0,\gamma}\xi^a\ .
\ee
By requiring (\ref{regime}), we must impose $c_a\xi^a_\mu/(M_2 M_{\rm P})\sim \sqrt{\varepsilon}$.
Having fixed the metric, the equation of motion for $\pi$ is clearly invariant under (\ref{ggf}).
However now, differently from the curved Galileon symmetry in section \ref{secII}, the symmetry (\ref{exact}) acts also on the metric $g_{\mu\nu}$, under which the equations should be approximately invariant by construction, if (\ref{regime}) and (\ref{HFregime}) are satisfied. This effect can be understood by observing that the  transformation for $g_{\mu\nu}$ can be regarded as an infinitesimal  $\pi$-dependent `diffeomorphism'
\be\label{diff}
g_{\mu\nu}\rightarrow g_{\mu\nu}+\nabla_{(\mu}w_{\nu)}\, , \qquad w_\mu=2\frac{c_a\xi^a_\mu \pi}{M_2^2 M_{\rm P}^2}\ .
\ee
We would like to end this section by commenting on ${\cal L}_3$. As discussed in \cite{fab}, 
\be
\int d^4x\sqrt{-g} {}^{**}R^{\alpha\beta\mu\nu}\pi_\alpha\pi_\mu\pi_{\beta\nu}&=&-\frac{1}{4}M_2^2M_{\rm P}^2\int d^4x \frac{\delta{\sqrt{-g}\pi GB}}{\delta g_{\alpha\beta}}\delta g_{\alpha\beta}\Big |_{\delta g_{\alpha\beta}=-\frac{\partial_\alpha\pi\partial_\beta\pi}{M_2^2M_{\rm P}^2 }}+{\rm boundaries}\ ,
\ee
where $GB$ is the Gauss-Bonnet combination.
Therefore, ${\cal L}_3$ might be also rewritten in terms of a Finsler metric in the high friction regime. However, the presence of the tadpole term $\pi GB$, would not be invariant under the symmetry (\ref{exact}).

We then focus in the rest of the paper only on the following $\pi$-parity invariant Lagrangian: 
\be\label{sloth2}
S_{\rm sloth}=\frac{1}{2}\int \d^4 x \sqrt{-g} \left[M_{\rm P}^2R-\left(g^{\alpha\beta}-\frac{G^{\alpha\beta}}{M^2}\right)\partial_\alpha\pi\partial_\beta\pi\right]\ ,
\ee
where we replaced $M_2\rightarrow M$ for notational simplicity.

\subsection{On the strong coupling in high friction regime}\label{strongsec}

 Because of the non-trivial coupling of gravity with the Slotheon, the identification of the tree-level strong coupling scale of the theory (\ref{sloth2}) is in general strongly background dependent. In other words, in order to calculate the perturbative cut-off scale of the theory (\ref{sloth2}) as an expansion of fields in specific backgrounds, one should take care upon identifying the correct propagating degrees of freedom which are generically a combination of the Slotheon, the graviton and the background quantities. 

In order to identify the physical perturbative degrees of freedom one should rewrite the theory as an expansion around gaussian fixed points. Although the obvious Minkowski cut-off of the theory (\ref{sloth2}) is $\Lambda_{\text{cut-off}}=(M^2 M_{\rm P})^{1/3}$, it has been shown in \cite{yuki, new} that for a slow rolling Slotheon in a homogeneous and isotropic background, (i.e. when $\varepsilon\ll 1$) the strong coupling scale of the theory is enhanced to $\Lambda_{\text{cut-off}}= M_{\rm P}+{\cal O}(\varepsilon)$. This result can be readily generalized by using the previous arguments. Indeed, let us consider an expansion around the $\varepsilon\ll 1$ solution, which we loosely call small derivative regime. 

If there exists a non-trivial background for the scalar field, one can always reparameterize time in order to reabsorb the perturbative scalar degree of freedom into the metric. Explicitly, let us consider the expansion of the Slotheon around a background solution $\pi_0$. At first order (higher orders are easily generalizable) we have
\be
\pi=\pi_0(t,\vec x)+\delta\pi(t,\vec x)\ ,
\ee
where $\delta\pi$ is the perturbation.

We can now consider the first order coordinate transformation $t\rightarrow t+\delta t$ to obtain, at first order
\be
\pi=\pi_0+\dot\pi_0\delta t+\delta\pi\ .
\ee
Therefore, by choosing the gauge $\delta t=-\delta\pi/\dot\pi_0$ we obtain the desired result of reabsorbing the scalar degree of freedom into the metric. This gauge is called unitary gauge in cosmology and widely used to calculate (quantum) correlation functions \cite{maldacena}.

Using the unitary gauge, during the small derivative regime of the background scalar field, the theory (\ref{sloth2}) is well approximated by (\ref{EH}) plus the canonical kinetic term, i.e.
\be\label{k3}
S=\frac{1}{2}\int d^4x \sqrt{-h}\left[M_{\rm P}^2R(h)-(\partial\pi)^2\right]\ .
\ee
Therefore, the true (gaussian) degrees of freedom become a self interacting $h_{\mu\nu}$ with cut-off scale $M_{\rm P}$ and the free scalar $\pi$. 

We thus proved that the strong coupling scale of the Slotheonic theory in a background in which  $\varepsilon\ll 1$ is
\be
\Lambda_{\text{cut-off}}= M_{\rm P}+{\cal O}(\varepsilon)\ ,
\ee
which matches direct computations in homogeneous and isotropic backgrounds \cite{yuki,new}. The presence of a possible (renormalizable) potential term for $\pi$ obviously does not alter this result.

In this sense then, the background $\varepsilon\ll 1$ is always in weak coupling if $M$ and curvatures are below the Planck scale.

%%%%%%%%%%%%%%%%%%%%%%%%%%%%%%%%%%%%%%%%%%%%%%%
%%%%%%%%%%%%%%%%%%%%%%%%%%%%%%%%%%%%%%%%%%%%%%%

\section{The Slotheon: a ``slow'' scalar field}

\noindent Let us now investigate the properties of  the theory (\ref{sloth2}).
In particular, we focus on the dynamics which governs the temporal evolution of the scalar field $\pi$. 

Let us take the ADM decomposition \cite{ADM,wald} where the metric can be written as
\be
\d s^2=-N^2\d t^2+\gamma_{ij}(\d x^i-N^i\d t)(\d x^j-N^j\d t)\ .
\ee
In this parameterization of the metric the action related to ${\cal L}_2$ looks like
\be
S=\frac{1}{2}\int \d^4 x\,N\sqrt{\gamma} \left[\left(\frac{1}{N^2}+\frac{G^{tt}}{M^2}\right)\dot \pi^2+2\left(\frac{N^i}{N}+\frac{G^{ti}}{M^2}\right)\dot\pi\partial_i\pi-(\gamma^{ij}-\frac{G^{ij}}{M^2})\partial_i\pi\partial_j\pi\right]\ .
\ee
The momentum conjugate to $\pi$ is therefore defined as
\be
\Pi=\frac{\delta S}{\delta \dot\pi}= N\sqrt{\gamma} \left[\left(\frac{1}{N^2}+\frac{G^{tt}}{M^2}\right)\dot \pi+\left(\frac{N^i}{N}+\frac{G^{ti}}{M^2}\right)\partial_i\pi\right]\ .
\ee
The contribution of the Hamiltonian density coming from the scalar field kinetic term is then
\be
{\cal K}_\pi=\frac{1}{2}\sqrt{\gamma}\,\alpha^2\frac{\dot \pi^2}{N^2}\ ,
\ee
where we defined
\be
\alpha^2\equiv 1+N^2\frac{G^{tt}}{M^2}\ .
\ee

We would now like to focus on the regimes in which $G^{tt}\geq 0$. This condition can be regarded as the analogous of standard weak energy condition in our non-canonical theory and immediately implies that 
\be
\alpha^2\geq 1\ .
\ee
Considering the same background geometry, we would like now to compare the kinetic energies of a canonical scalar field and of the Slotheon. In order to do that we must fix the time lapse to be the same for the two theories. The simplest choice is to use the synchronous gauge $N=1$. In this case, for a given kinetic energy per unit volume (${\cal K}_\pi$) we have
\be
\dot\pi^2\sim\frac{{\cal K}_\pi}{\alpha^2}\leq {\cal K}_\pi\ .
\ee
It is then clear that the time derivative of the Slotheon is smaller than the corresponding one (with the same energy density) of a canonical scalar field (where $\alpha=1$). In this sense the Slotheon is slower than a canonical scalar field. 

The same conclusion can be readily drawn also by adding a positive definite potential. Although we imposed the $\pi$-parity invariance to select the theory all our results in this section and in the following sections are also valid for potentials breaking this invariance. 

The Slotheonic theory is then 
\be\label{withV}
\tilde S=\frac{1}{2}\int \d^4x \sqrt{-g}\left[M_{\rm P}^2 R-\left(g^{\mu\nu}-\frac{G^{\mu\nu}}{M^2}\right)\pi_{\mu}\pi_{\nu}-2V(\pi)\right]\ ,
\ee
with $V(\pi)\geq 0$ and it is easy to see that this modification does not modify the above arguments. Notice also that the slowing of the Slotheon is due solely to gravitational interaction. This is profoundly different from self-interacting theories which have similar properties {\it only} in specific backgrounds (see for instance \cite{selfinteraction}).
 
A typical example of a Slotheonic theory in action can be seen on de Sitter or almost-de Sitter (inflationary) backgrounds, where $G^{tt}=3\Lambda^2$ and $\Lambda$ is roughly constant. In this case the scalar field kinetic energy of a canonical scalar field is modified as
\be
\dot\pi^2\rightarrow(1+3\frac{\Lambda^2}{M^2})\dot \pi^2\ .
\ee
Redefining the effective time of the scalar field as
\be
\d t_{\rm Slotheon}=\frac{\d t}{\sqrt{1+3\frac{\Lambda^2}{M^2}}}\ ,
\ee
we find
\be
\d t_{\rm Slotheon}\leq \d t\ .
\ee
Therefore, one may interpret the proper clock of the Slotheon to be slower than the clock of an observer tight to the Universe expansion. From a different point of view, the slowness of the Slotheon in the previous example was obtained by increasing the friction term acting on the scalar field.  This mechanism has been dubbed the Gravitational-Enhanced-Friction mechanism in \cite{yuki} for inflationary scenarios and it is the base of New Higgs Inflation \cite{new} and UV-Protected inflation \cite{uv,yuki}.

\section{No-(Slotheonic) Hair Theorem}\label{no-hair}

\noindent In this section we will prove that the only spherically symmetric black hole solution of the Slotheonic theory (\ref{withV})
 is the vacuum solution, i.e.\ the Schwarzschild solution. This result is an important step to prove the stability of the Slotheonic theory in general curved space-time (which we postpone for future work). In fact, it is widely believed that ghost-like or unstable scalar theories may support scalar hairs outside a black hole horizon \cite{bekenstein95}.
 
Since we assumed that the potential does not violate energy conditions, we can restrict our proof to the massless case. In fact, since a mass implies a faster decay of the scalar field than the massless case, proving the impossibility of massless scalar hairs will be enough. We will then restrict our attention to the theory
\be\label{SS}
S=\frac{1}{2}\int \d^4x \sqrt{-g}\left[M_{\rm P}^2\, R-\left(g^{\mu\nu}-\frac{G^{\mu\nu}}{M^2}\right)\pi_{\mu}\pi_{\nu}\right]\ .
\ee
In order to prove that the only spherically symmetric solution is trivial for the Slotheon we will closely follow \cite{bekenstein} with the help of the gravity and scalar field equations obtained by
varying the action (\ref{SS}) with respect to $\pi$ and metric. The equations are respectively (see also \cite{shush})
\ba
&&(g^{\mu\nu}-\frac{G^{\mu\nu}}{M^2})\pi_{\mu\nu}=0,\nonumber
 \\
 \\
&&G_{\mu\nu}=M_{\rm P}^{-2} T_{\mu\nu}\ .\nonumber
\ea
Where
\be\label{T}
T_{\mu\nu}=\pi_\mu\pi_\nu-\frac{1}{2}g_{\mu\nu}(\partial\pi)^2+\frac{\Theta_{\mu\nu}}{M^2}\ ,
\ee
and
\be
\Theta_{\mu\nu}=\frac{1}{2}\pi_{\mu}\pi_{\nu}R-2\pi_{\alpha}\pi_{(\mu}R^{\alpha}_{\nu)}+\frac{1}{2}\pi_{\alpha}\pi^{\alpha}G_{\mu\nu}-\pi^{\alpha}\pi^{\beta}R_{\mu\alpha\nu\beta}-\pi_{\alpha\mu}\pi^{\alpha}_\nu+\pi_{\mu\nu}\pi_\alpha^{~\alpha}+\frac{1}{2}g_{\mu\nu}[\pi_{\alpha\beta}\pi^{\alpha\beta}-(\pi_\alpha^{~\alpha})^2+2\pi_\alpha\pi_\beta R^{\alpha\beta}]\ .\nonumber
\ee

\subsection{Spherically symmetric case}

\noindent Let us start by imposing spherical symmetry. In this case the metric will be
\be\label{ds}
\d s^2=-A(r)^2\d t^2+B(r)^2\d r^2+r^2\d\Omega^2\ ,
\ee
where $d\Omega^2=\d\theta^2+\sin\theta^2\d\phi^2$.

\begin{figure}
  \begin{center}
    \includegraphics[width=0.5\textwidth]{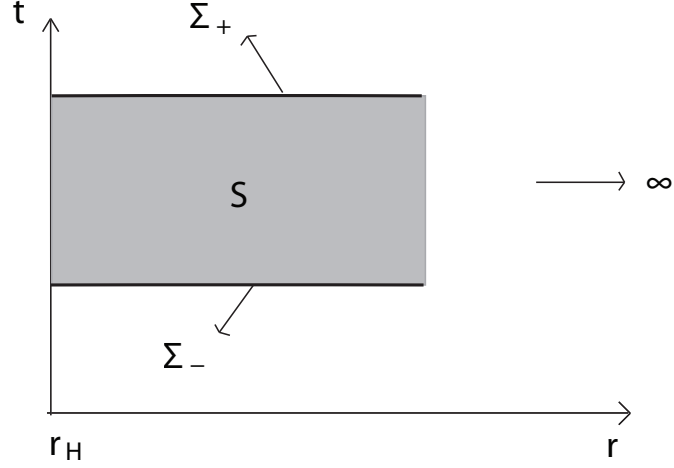}
 \end{center}
  \caption{Integration region.}\label{sloth}
     \end{figure}

The equation of motion for the scalar field reads
\be
\left(g^{\alpha\beta}-\frac{G^{\alpha\beta}}{M^2}\right)\nabla_\alpha\nabla_\beta\pi=0\ .
\ee
Multiplying it by the scalar field $\pi$ and integrating in the closed region $S$ of Fig.\ref{sloth}, delimited by an horizon at $r_H$ and two time slices $\Sigma_{\pm}$, we get
\be\label{int}
\int_S \d^4x \sqrt{-g}\pi\left(g^{\alpha\beta}-\frac{G^{\alpha\beta}}{M^2}\right)\nabla_\alpha\nabla_\beta\pi=0\ .
\ee
     
Integrating by parts (\ref{int}) we obtain
\be\label{int2}
\int_S \d^4x \sqrt{-g}\left(g^{\alpha\beta}-\frac{G^{\alpha\beta}}{M^2}\right)\nabla_\alpha\pi\nabla_\beta\pi=\int_H \d^3x \sqrt{-g}n_{\alpha}\left(g^{r\alpha}-\frac{G^{r\alpha}}{M^2}\right)\pi\pi'\ ,
\ee
where the sum of the boundary integrals over $\Sigma_{\pm}$ vanish because of staticity and the integral at infinity vanishes because the assumption of asymptotic flatness. In (\ref{int2}) $H$ is the horizon surface, $n^{\alpha}$ is the normal to the horizon and ${}'=d/dr$. By definition an horizon is a light-like surface, i.e. $n_\alpha n^\alpha=0$ and for a static metric $n_t=0$ on the horizon. By using the Cauchy inequality
\be
0\leq (n_{i}A^{i})^2\leq n_i n^i A_j A^j=0\ ,
\ee
where the last equality is valid if and only if $A_j A^{j}<\infty$, we find that the left hand side of (\ref{int2}) vanishes. Taking
\be
A^j=\left(g^{rj}-\frac{G^{rj}}{M^2}\right)\pi\pi'\ ,
\ee
we see indeed that $A_j A^j$ cannot diverge for a smooth space-time and non-divergent scalar field. We are then left with the integral equation
\be\label{lo}
\int_S \d^4x \sqrt{-g}\left(g^{\alpha\beta}-\frac{G^{\alpha\beta}}{M^2}\right)\nabla_\alpha\pi\nabla_\beta\pi=\int_S \d^4x \sqrt{-g}\left(g^{rr}-\frac{G^{rr}}{M^2}\right)\pi'^2=0\ .
\ee
We are now interested in finding the form of $G^{rr}$. The gravity equations are (we fix here $M_{\rm P}$=1)
\be
G_{\alpha\beta}=T_{\alpha\beta}\ ,
\ee
where $T_{\alpha\beta}$ is given in Eq. (\ref{T}). With the metric (\ref{ds}) we find
\be
G_{rr}=\frac{\frac{1}{2}-\frac{1}{M^2r^2}}{1+\frac{3}{2}\frac{\pi'^2}{B^2M^2}}\pi'^2\ .
\ee
Plugging the previous result into the integral (\ref{lo}) we get
\be\label{41}
\int_S \d^4x \sqrt{-g}\frac{B^2+\frac{\pi'^2}{M^2}(1+\frac{1}{r^2})}{B^4(1+\frac{3}{2}\frac{\pi'^2}{B^2 M^2})}\pi'^2=0\ .
\ee
Since the integrand is positive definite in (\ref{41}), the only solution is $\pi'=0$, i.e. the only solution for a spherically symmetric black hole is with no Slotheonic hairs. The black hole solution is then a solution of the Einstein equation in vacuum that has as the {\it unique} solution the Schwarzschild metric
\be
\d s^2=-(1-\frac{2m}{r})\d t^2+(1-\frac{2m}{r})^{-1}\d r^2+r^2\d\Omega^2\ ,
\ee
where $m$ is the black hole mass.

\subsection{No-hair theorem: a re-interpretation and a conjecture}

\noindent We can now re-interpret the no-hair theorem proved previously in the theory (\ref{sloth2}). If we consider the canonical theory (\ref{k3}) we can obviously  use the standard no-hair theorem. In that case there is no non-trivial solution for the scalar field $\pi$ and the only spherically symmetric solution is the Schwarzschild solution
\be
\d s^2=-(1-\frac{2m}{r})\d t^2+(1-\frac{2m}{r})^{-1}\d r^2+r^2\d\Omega^2\ .
\ee
Of course, the theories (\ref{sloth2}) and (\ref{k3}) are equivalent only up to fist order in the small derivative   perturbative regime (\ref{regime}).
Hence, the no-hair theorem for the canonical theory (\ref{k3}) can only be used to easily conclude that black hole solutions in (\ref{sloth2}) cannot have {\it perturbative} Slotheonic hairs. This provides a non-trivial confirmation of our direct proof of the no-hair theorem in Sec.\ref{no-hair} for spherically symmetric black holes and it automatically extends to non-spherically symmetric black holes, within the perturbative regime. This encourages us to {\it conjecture} that there are no black hole solutions with non-perturbative Slotheonic hairs. We leave the investigation of this important conjecture for future work.

\section{Asymptotic local shift symmetry and inflation}

\noindent Standard inflationary models enjoy an asymptotic shift symmetry of the scalar \cite{creminelli}
\be\label{sym}
\pi\rightarrow\pi+c\ ,
\ee
due to the fact that, under such a shift, the Inflaton potential only shifts at next to leading order in the slow roll expansion. This shift however, does not protect the theory under new derivative couplings and it is expected to be anyway broken by Quantum Gravity effects. In order to avoid these potential problems one may then try to ``gauge'' the symmetry (\ref{sym}) to
\be\label{sym2}
\pi\rightarrow\pi+f(x)\ .
\ee
In a spatially flat Friedamn-Robertson-Walker (FRW) geometry
\be
\d s^2=-\d t^2+a(t)^2 \d\vec {x}\cdot \d\vec{x}\ ,
\ee
the field and gravity evolution equations are \cite{new}
\be\label{eq0}
&&H^2=\frac{1}{3M_{\rm P}^2}\left[\frac{\dot\pi^2}{2}(1+9\frac{H^2}{M^2})+V\right],\cr
&&\partial_t\left[a^3\dot\pi(1+3\frac{H^2}{M^2})\right]=-a^3V'\ ,
\ee
where $H=\frac{\dot a}{a}$ and $(\dot{})=\d/\d t$.

In GEF of \cite{yuki,new,uv}, the Inflaton (a Slotheon) is non-minimally coupled to gravity as in (\ref{sloth2}) so that slow roll may be naturally obtained. With this coupling, even very steep potentials for the scalar field, $V(\pi)$, would produce a successful inflationary scenario, thanks to a huge gravitational friction acting on the Inflaton during inflation.
Specifically, one can then always choose the mass $M$ small enough such that, during inflation, $H^2/M^2\gg 1$. Note, as explained before, that no strong coupling happens here thanks to the canonical normalization of the field $\pi$ \cite{new, yuki}. This regime is called the high friction regime \cite{yuki}. In this regime, for any given potential $V$, a quasi-de Sitter solution always exists for $M$ small enough. This is the basis for the New Higgs Inflation \cite{new} and the UV-protected Inflation \cite{uv}. A quasi-de Sitter background implies that the slow roll parameters are small, i.e.
\be\label{high}
\epsilon\equiv-\frac{\dot H}{H^2}\ll 1\ ,\ \delta\equiv\Big|\frac{\ddot\pi}{H\dot\pi}\Big|\ll 1\ .
\ee

We will firstly focus on the case in which the Inflaton potential has small curvatures (chaotic type inflation), at least during inflation. We then ask that the ``canonical'' slow roll conditions are satisfied 
\be\label{grsr}
\epsilon_{\rm can}\equiv\frac{V'^2}{2V^2}M_{\rm P}^2\ll 1\ ,\ \eta_{\rm can}\equiv\frac{V''}{V}M_{\rm P}^2\ll1\ ,
\ee
and assume a monomial potential for the Inflaton so that the above conditions generically require $\pi\gg M_{\rm P}$. 

In high friction limit ($H\gg M$), during slow roll, one finds that \cite{yuki}
\be\label{smallexp}
\epsilon\simeq\frac{3}{2}\frac{\dot\pi^2}{M^2M_{\rm P}^2}\simeq\epsilon_{\rm can}\frac{M^2}{3H^2}\ll 1\ ,\ V=V_0\left[1+{\cal O}(\sqrt{\epsilon})\frac{\delta\pi}{M_{\rm P}}\right]\ ,
\ee
where $\delta\pi$ is a shift on the background value for $\pi$ and $\epsilon$ is the true slow roll parameter defined in (\ref{high}). Thus, it is exactly in this regime that the symmetry (\ref{exact}) is realized for the kinetic and gravitational parts of the action as, in this regime, $\varepsilon\sim\epsilon\ll1$. The potential term also would break the symmetry (\ref{exact}) only at higher order in slow roll. This can be easily seen from the action. There, the potential term would shift as
\be
V\sqrt{-g}\rightarrow V\sqrt{-g}\left(1+{\cal O}(\sqrt{\epsilon})\frac{f}{M_{\rm P}}\right)\ ,
\ee
thanks to (\ref{smallexp}). In other words, the local shift symmetry (\ref{exact}) is only softly broken by the potential if (\ref{grsr}) are satisfied.

Let us now suppose that the potential generating inflation violates the conditions (\ref{grsr}). For monomial potential this would mean sub-Planckian field values. The GEF mechanism would nevertheless work in order to fulfill (\ref{high}) for $M$ small enough. This can be easily seen from the first equation in (\ref{smallexp}), which is always valid in high friction limit \cite{yuki}. In this case the symmetry (\ref{exact}) would in general be badly broken by the potential, unless the potential does not introduce any self interactions. In other words, the symmetry (\ref{exact})  may still be softly broken by a mass term, i.e.\ in the case in which $V=V_0\pm\frac{1}{2}m^2\pi^2$, for {\it any} field value during inflation. Because no self-interactions are introduced in the potential, one would indeed expect that quantum corrections to the propagator would still be suppressed by slow roll, i.e.\ they would still obey the asymptotic symmetry (\ref{exact}) during inflation \footnote{For a similar discussion see also \cite{galinf}.}. Thanks to that, the UV-protected inflation of \cite{uv,yuki}, has an extra quantum protection in the high friction limit: the local shift symmetry (\ref{exact}).

We then found that during inflation and in high friction regime, the Slotheonic Lagrangian (\ref{withV}) enjoys an asymptotic gauge symmetry (\ref{exact})  protecting chaotic type inflationary set-up and inflationary set-up with mass potentials from quantum corrections to both the potential and the kinetic terms. Note that extra-derivative couplings that could be added and are invariant under the approximate symmetry (\ref{exact}), can only come from further expanding the action (\ref{einstein}). Therefore, extra-derivative couplings may only modify the equation to higher order in slow roll, in this sense they are completely negligible and the inflationary trajectory is stable. One may still wonder about couplings of the Slotheon to matter fields. These couplings would generically produce a (Coleman-Weinberg) logarithmically corrected potential for the effective canonically normalized field $\psi=\frac{H}{M}\pi$. The strength of these corrections depends upon the particular couplings chosen and may or may not be important for the Inflationary evolution. We leave this important discussion for a future work.

Finally, let us comment on possible renormalization of the mass parameter suppressing the non-minimal coupling ${\cal L}_2^{\rm nm}$. Since, during the high friction regime, the graviton is still canonically normalized with the Planck scale (see Sec. \ref{strongsec}), we expect that the running of $M$ is suppressed by the Planck scale and therefore negligible for an inflationary trajectory where the total energy is far below $M_{\rm P}$. The study of the exact running of $M$ is left for future work.

\section*{Acknowledgements}
We wish to thank Alex Vikman for careful reading and important comments on the first draft of the paper. CG wishes to thank Paolo Creminelli and Guido D'Amico, for important comments. CG and PM would also like to thank Yuki Watanabe for useful discussions. PM wishes to thanks Claudia de Rham for comments on the first draft. The Authors thanks Alex Kehagias for earlier participation to the project and important discussions. Finally, the Authors wish to thank the anonymous referee for pointing out an extra Galilean invariant term. CG and PM are supported by Alexander Von Humboldt Foundation. LM is partially supported by the ERC Advanced Grant n.226455 ``Superfields", by the Italian MIUR-PRIN contract 20075ATT78 and by the NATO grant PST.CLG.978785.

 \end{document}